\documentclass[11pt]{article}

\usepackage{acl}
\usepackage{amsmath}
\usepackage{amssymb}

\usepackage{fontspec}
\usepackage{xeCJK}

\setCJKsansfont{FandolHei-Regular.otf}
\setCJKmonofont{FandolFang-Regular.otf}

\usepackage{xurl}
\usepackage{microtype}
\usepackage{graphicx}

\usepackage{booktabs}
\usepackage{multirow}
\usepackage{tabularx}
\usepackage{makecell}
\usepackage{array}
\usepackage{float}

\usepackage{tikz}
\usepackage{pgf-pie}

\usepackage[most]{tcolorbox}
\usepackage{fvextra}

\newtcolorbox{promptbox}[1][]{
    enhanced,
    breakable,
    colback=gray!8,
    colframe=gray!35,
    boxrule=0.5pt,
    arc=2pt,
    left=6pt,
    right=6pt,
    top=6pt,
    bottom=6pt,
    #1
}

\author{
\textbf{Yunfei Zhong$^{1*}$ \quad
Jun Yang$^{1*}$ \quad
Wei Huang$^{1}$ \quad
Yinqiong Cai$^{2\dagger}$ \quad
Haosheng Qian$^{2}$} \\
\textbf{Yixing Fan$^{1}$ \quad
Ruqing Zhang$^{1}$ \quad
Lixin Su$^{2\dagger}$ \quad
Daiting Shi$^{2}$ \quad
Jiafeng Guo$^{1}$} \\
$^{1}$State Key Laboratory of AI Safety, Institute of Computing Technology, Chinese Academy of Sciences \\
$^{1}$University of Chinese Academy of Sciences, Beijing, China \\
$^{2}$Baidu Inc., Beijing, China \\
\texttt{\{zhongyunfei25s,yangjun24s,huangwei21b\}@ict.ac.cn} \\
\texttt{\{fanyixing,zhangruqing,guojiafeng\}@ict.ac.cn} \\
\texttt{\{caiyinqiong,qianhaosheng,sulixin,shidaiting01\}@baidu.com}
}

\title{Querit-Reranker: Training Compact Multilingual Rerankers via Efficient Label-Free Distribution Adaptation}

\begin{document}
\maketitle

\begingroup
\renewcommand{\thefootnote}{\fnsymbol{footnote}}
\footnotetext[1]{Co-first authors.}
\footnotetext[2]{Corresponding authors.}
\endgroup

\begin{abstract}
A deployable multilingual reranker must not only generalize across languages, domains, and ranking tasks, but also remain efficient to serve as a second-stage reranker in practical systems. However, adapting it to new target distributions typically requires extensive task-specific relevance annotations. We present \textbf{Querit-Reranker}, a family of multilingual rerankers trained with a data-centric pipeline for label-efficient adaptation. We instantiate it as \textbf{Querit-Reranker-A0.4B}, initialized from an in-house MoE backbone with 0.4B activated parameters, and \textbf{Querit-Reranker-4B}, initialized from Qwen3-Embedding-4B. Our pipeline first learns general relevance modeling from large-scale ranking-oriented data, then adapts to target distributions through synthetic-query mining with teacher scores as continuous soft labels. To consolidate complementary task-adapted strengths, we further merge checkpoints via spherical linear interpolation, obtaining a single deployable model without runtime ensembling overhead. Using Qwen3-Embedding-0.6B as the first-stage retriever, \textbf{Querit-Reranker-A0.4B} improves average nDCG@10 from 54.11 to 59.28 on BEIR and from 59.87 to 67.70 on MIRACL. On MTEB Multilingual v2 Reranking, it also shows competitive performance with larger embedding-based models, while \textbf{Querit-Reranker-4B} further achieves state-of-the-art performance among all evaluated models. We release both models on Hugging Face.\footnote{\href{https://huggingface.co/Querit/Querit}{Querit-Reranker-A0.4B} and \href{https://huggingface.co/Querit/Querit-4B}{Querit-Reranker-4B}.}

\end{abstract}

\section{Introduction}
Modern information access systems, such as AI-driven retrieval-augmented generation (RAG) architectures \cite{gao2024retrievalaugmentedgenerationlargelanguage}, commonly rely on a retrieve-then-rerank pipeline to balance efficiency and accuracy. 
An efficient first-stage retriever identifies candidate documents, which are then re-ranked by a more sophisticated second-stage reranker for enhanced precision. 
Rerankers for multilingual retrieval must balance fine-grained relevance across languages, domains, and tasks with the efficiency for second-stage deployment.

There are two challenges for multilingual rerankers. One is the backbone-efficiency trade-off. Strong multilingual backbones have become the de facto standard for retrieval and reranking models \cite{chen2024bgem3,zhang2024mgte,zhang2025qwen3embeddingadvancingtext}. Yet their effectiveness often comes at the cost of inference efficiency, and striking a practical balance remains an open challenge.
The other is the adaptation-deployment gap. Task-specific fine-tuning can yield complementary gains, but deploying multiple adapted models, or runtime ensembles thereof, would linearly scale up latency and memory, making such strategies impractical for real-world serving. This tension is further amplified in multilingual scenarios, where per-task annotations are costly and target distributions vary widely.

These motivate our central question: how can we adapt multilingual backbones into accurate and efficient rerankers with limited task-specific human annotation? We address this question with \textbf{Querit-Reranker}, a family of multilingual cross-encoder rerankers trained through a data-centric adaptation pipeline. We instantiate the pipeline in two settings: \textbf{Querit-Reranker-A0.4B}, initialized from Querit-A0.4B, an in-house MoE backbone with 0.4B activated parameters, and \textbf{Querit-Reranker-4B}, initialized from Qwen3-Embedding-4B\footnote{\href{https://huggingface.co/Qwen/Qwen3-Embedding-4B}{Qwen3-Embedding-4B}}. Both models follow the standard cross-encoder paradigm: each query-document pair is jointly encoded and mapped to a scalar relevance score, making the models directly compatible with retrieve-then-rerank pipelines.

\begin{figure}[t]
\centering
\includegraphics[width=\linewidth]{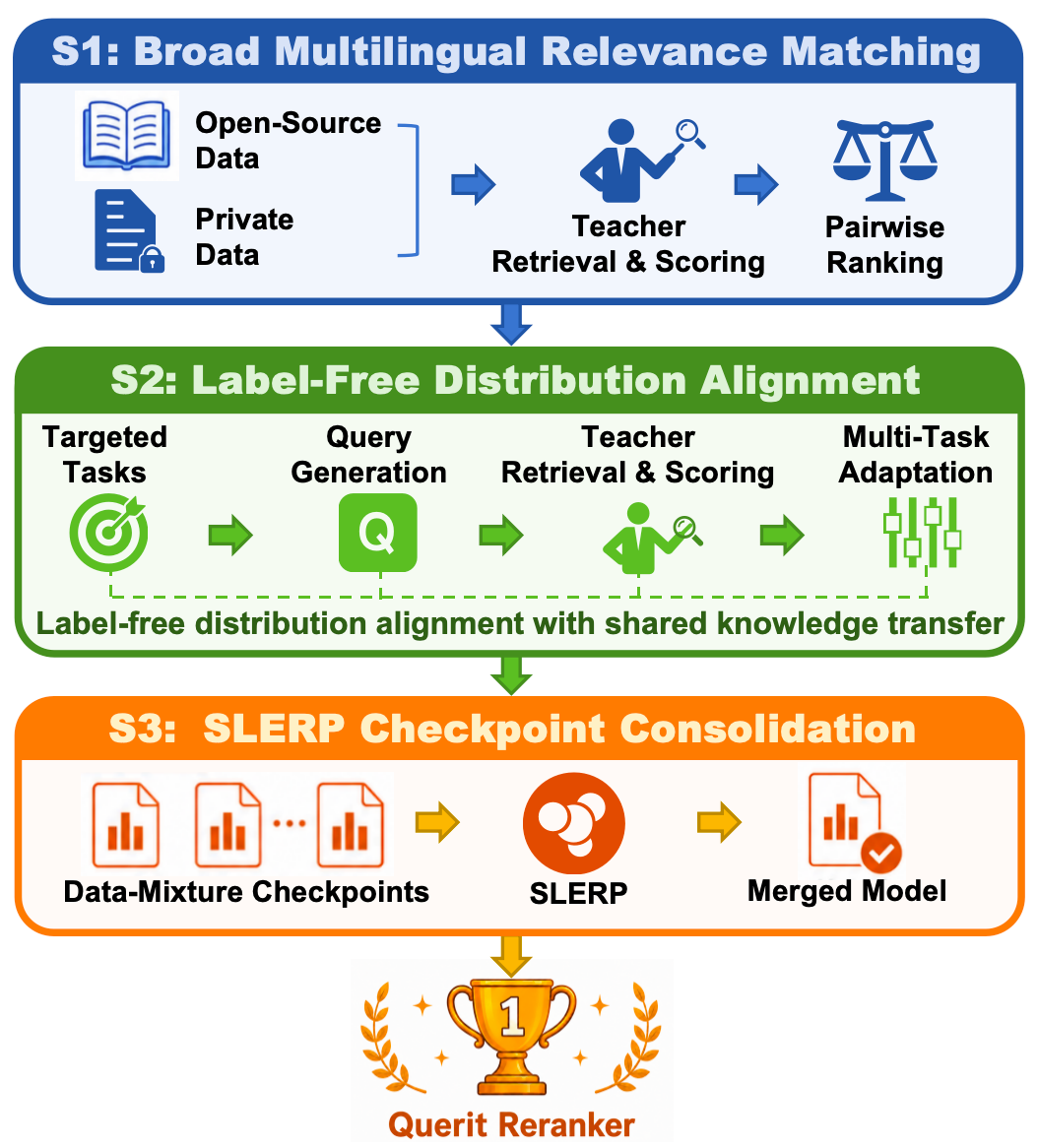}
\caption{Training pipeline of Querit-Reranker.}
\label{fig:querit-reranker-pipeline}
\end{figure}

As illustrated in Figure~\ref{fig:querit-reranker-pipeline}, our training pipeline consists of three stages. First, we train on large-scale ranking-oriented data, including open-source retrieval and reranking data as well as high-quality de-identified private data, to learn broad multilingual relevance modeling. Second, we perform targeted synthetic-query mining on representative MMTEB reranking corpora. Document-conditioned synthetic queries are paired with teacher scores as continuous soft relevance labels, enabling label-efficient adaptation to target ranking distributions without requiring additional dense human relevance annotations for the target corpora. Third, we merge selected checkpoints with spherical linear interpolation (SLERP), consolidating complementary task-adapted strengths into a single deployable reranker without runtime ensembling overhead.

We evaluate Querit-Reranker across multiple benchmarks: MMTEB reranking tasks from MTEB-multilingual-v2 and MTEB-eng-v2 \cite{muennighoff2022mteb, enevoldsen2025mmtebmassivemultilingualtext}, BEIR for general English retrieval \cite{thakur2021beirheterogenousbenchmarkzeroshot}, and MIRACL for multilingual retrieval \cite{zhang2023miracl}. Using Qwen3-Embedding-0.6B as the shared first-stage retriever, \textbf{Querit-Reranker-A0.4B} improves average nDCG@10 from 54.11 to 59.28 on BEIR and from 59.87 to 67.70 on MIRACL. \textbf{Querit-Reranker-4B} further achieves the highest average score among publicly available models on the MTEB-multilingual-v2 reranking leaderboard as of August 29, 2026.

Our contributions are summarized as follows:
\begin{itemize}
\item We introduce \textbf{Querit-Reranker}, a family of multilingual cross-encoder rerankers, including a compact MoE-based variant and a larger dense-backbone variant, both designed for standard retrieve-then-rerank deployment.
\item We present an integrated, deployment-oriented data-centric training pipeline for label-efficient target adaptation: large-scale ranking-oriented training builds general relevance modeling, synthetic-query mining with teacher scores adapts the model to target ranking distributions, and SLERP-based checkpoint merging consolidates complementary strengths without runtime ensembling.
\item We evaluate Querit-Reranker across multiple benchmarks, including MMTEB, BEIR, and MIRACL, showing that Querit-Reranker-A0.4B achieves strong parameter efficiency and that Querit-Reranker-4B reaches state-of-the-art performance among publicly available multilingual reranking models.

\end{itemize}

\section{Related Work}
\label{sec:related-work}

\paragraph{Neural ranking architectures.}
Neural ranking models are commonly divided into representation-based and interaction-based methods \cite{guo2020deep}. Representation-based methods encode queries and documents separately, enabling efficient candidate scoring but limiting explicit token-level interaction \cite{reimers2019sentencebertsentenceembeddingsusing,karpukhin2020densepassageretrievalopendomain}. Interaction-based rerankers, especially cross-encoders, jointly encode query-document pairs and are therefore well suited for second-stage reranking over a limited candidate set \cite{nogueira2019passagererankingbert,nogueira2019multistagedocumentrankingbert,macavaney2019cedr}. Late-interaction models such as ColBERT provide a middle ground by preserving token-level matching with lower retrieval-time cost \cite{khattab2020colbertefficienteffectivepassage}. Querit-Reranker follows the cross-encoder paradigm to retain fine-grained query-document interaction while targeting efficient second-stage deployment.

\paragraph{Multilingual retrieval and reranking with strong backbones.}
Recent multilingual embedding models have shown that large pretrained backbones, multilingual supervision, and synthetic or distilled training signals can substantially improve retrieval quality. Representative systems include E5-Mistral, BGE-M3, Gemini Embedding, Llama-Embed-Nemotron-8B, jina-embeddings-v5, GTE, KaLM-Embedding-V2, and F2LLM-v2 \cite{wang2024e5mistral,chen2024bgem3,lee2025geminiembedding,babakhin2025llamaembednemotron8buniversaltextembedding,akram2026jinaembeddingsv5texttasktargetedembeddingdistillation,li2023towards,zhao2025kalmembeddingv2,zhang2026f2llmv2inclusiveperformantefficient}. Multilingual rerankers such as mGTE, Qwen3-Reranker, jina-reranker-v3, and RankLLaMA further demonstrate the importance of reranking-oriented post-training for converting strong backbones into effective relevance scorers \cite{zhang2024mgte,zhang2025qwen3embeddingadvancingtext,wang2025jinarerankerv3lateinteractionlistwise,ma2023rankllama}. Querit-Reranker builds on this line, but focuses on label-efficient target adaptation for deployable multilingual reranking.

\paragraph{Synthetic supervision and model merging.}
When task-specific relevance labels are limited, retrieval systems can construct additional supervision through generated queries and teacher-provided scores. Prior work has explored LLM-based synthetic query generation and teacher distillation for retrieval training \cite{bonifacio2022inparsdataaugmentationinformation,dai2022promptagatorfewshotdenseretrieval,jeronymo2023inparsv2largelanguagemodels,hofstatter2020marginmse}. In parallel, parameter-space merging methods combine multiple checkpoints without increasing inference-time cost \cite{wortsman2022modelsoups,ilharco2023editingmodelstaskarithmetic,yadav2023ties,goddard2024arceesmergekittoolkitmerging}. Querit-Reranker organizes these components into a staged recipe for multilingual cross-encoder reranking: synthetic-query mining with teacher scores adapts the model to target ranking distributions, and SLERP-based merging consolidates complementary strengths into a single deployable reranker.

\section{Model Architecture}
\label{sec:model-architecture}

Querit-Reranker adopts a cross-encoder scoring formulation for text reranking.
Given a query $q$ and a candidate document $d$, the model encodes the query-document pair in a single forward pass and outputs a scalar relevance score for second-stage ranking.
This formulation can be instantiated with different foundation backbones: Querit-Reranker-A0.4B is initialized from Querit-A0.4B, a self-developed MoE foundation model, while Querit-Reranker-4B is initialized from Qwen3-Embedding-4B.
Unless otherwise specified, we describe the shared reranking formulation below in a backbone-agnostic manner.

% \begin{figure}[t]
%     \centering
%     \includegraphics[width=\linewidth]{figures/Querit-Reranker-Architecture-v2.png}
%     \caption{Architecture of Querit-Reranker.}
%     \label{fig:querit-reranker-architecture}
% \end{figure}

For each query-document pair, Querit-Reranker formats the input as a relevance judgment prompt:
\begin{equation}
\mathbf{X} = [\mathrm{Inst}; q; t; c; [\mathrm{CLS}]],
\end{equation}
where $\mathrm{Inst}$ denotes the task instruction, $q$ the query, and $t$ and $c$ the document title and content, respectively.
The final $[\mathrm{CLS}]$ token is used as the aggregation position for relevance estimation.
The backbone encodes the input as $\mathbf{H}=[\mathbf{h}_1,\ldots,\mathbf{h}_n]=\mathrm{Backbone}_{\theta}(\mathbf{X})$. Since the appended $[\mathrm{CLS}]$ token occupies the last position, we use $\mathbf{h}_n$ as the relevance representation and feed it into a lightweight binary classification head:
\begin{equation}
\mathbf{p}=\mathrm{softmax}(\mathbf{W}\mathbf{h}_n+\mathbf{b})=[p_0,p_1].
\end{equation}
Here, $p_0$ and $p_1$ denote low- and high-relevance probabilities, and the final ranking score is computed as their expectation over $[-1,1]$:
\begin{equation}
    s(q,d) = [-1,1]\cdot \mathbf{p}^{\top} = -p_0 + p_1.
\end{equation}
A larger score indicates stronger relevance between the query and the candidate document.

\section{Model Training}
\label{sec:model-training}

We train Querit-Reranker with a unified three-stage reranking recipe.
Stage I builds broad multilingual query-document relevance matching ability from open-source retrieval and reranking datasets together with real-world business-scenario data.
Stage II performs label-free adaptation toward target ranking distributions through synthetic-query mining and unified teacher scoring, while Stage III consolidates complementary checkpoints from different data mixtures and training runs into a single deployable reranker.

\subsection{Training Objective}
Given a query--document pair $(q,d)$ with continuous relevance label $y_{q,d}$, Querit-Reranker outputs a scalar score $s(q,d)$. We optimize a pairwise hinge loss over documents associated with the same query. For each ordered pair $(d_i,d_j)$ satisfying $y_{q,d_i}>y_{q,d_j}$, the model is encouraged to assign a higher score to $d_i$:
\begin{equation}
\begin{aligned}
\mathcal{L}_{\mathrm{rank}}
&=
\frac{1}{|\mathcal{P}|}
\sum_{(i,j)\in\mathcal{P}} \\
&\quad
\max\!\left(0,\,m(y_i-y_j)-(s_i-s_j)\right),
\end{aligned}
\label{eq:rank_loss}
\end{equation}
where $\mathcal{P}$ denotes the set of such document pairs, $s_i=s(q,d_i)$, $y_i=y_{q,d_i}$, and $m>0$ is a margin coefficient. The margin is scaled by the teacher-score gap, enforcing larger score differences for pairs with larger relevance gaps.

\subsection{Stage I: Large-scale High-quality Data Collection}
\begin{figure}[t]
    \centering
    \scalebox{1}[1]{%
        \includegraphics[width=0.95\linewidth]{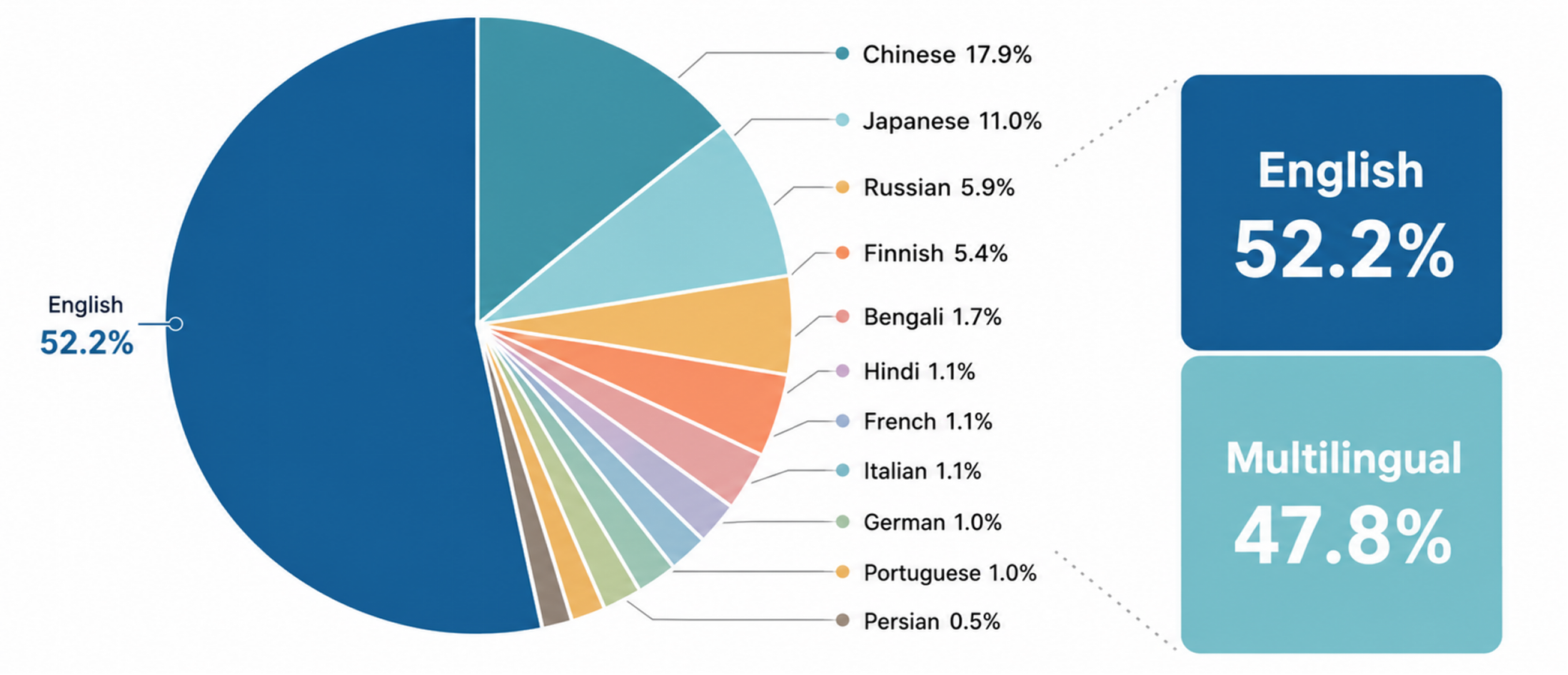}
    }
    \caption{Language distribution of Stage I training data.}
    \label{fig:open-source-lang}
\end{figure}

The first stage constructs broad reranking supervision from open-source retrieval and reranking datasets together with private high-quality search data from real-world business scenarios, which is de-identified before use and does not contain user-identifiable information.
The overall data composition and scale are summarized in Appendix~\ref{app:training-data-statistics}, Table~\ref{tab:synthetic-data-statistics}.
For the open-source portion, the selected datasets cover web search, multilingual retrieval, question answering, duplicate question detection, and news recommendation.
Its language distribution is shown in Figure~\ref{fig:open-source-lang}, with English as the largest portion and multiple non-English languages providing multilingual supervision.

For each query, we use Llama-Embed-Nemotron-8B\footnote{\href{https://huggingface.co/nvidia/llama-embed-nemotron-8b}
{Llama-Embed-Nemotron-8B}} to retrieve and score the top-100 candidates from the corresponding corpus, treating embedding similarity as continuous relevance labels.
To preserve relevance diversity, we partition the teacher-score range into 16 equal-width intervals and sample one candidate from each non-empty interval, filling missing slots with unused candidates closest to the interval centers.
The selected candidates are sorted by teacher score for optimization with Eq.~\eqref{eq:rank_loss}.

\subsection{Stage II: Targeted Synthetic-data Mining}
After Stage I, several MMTEB reranking tasks still show clear improvement space compared with leading leaderboard models.
We therefore perform targeted synthetic-data mining to adapt the reranker to target corpus distributions and query styles.
Specifically, we select four representative multilingual reranking tasks: AlloprofReranking, RuBQReranking, T2Reranking, and VoyageMMarcoReranking.
These tasks cover French, Russian, Chinese, and Japanese across educational question answering, factoid retrieval, noisy web-style retrieval, and MS MARCO-style search.

For each corpus, we sample documents and use the DeepSeek-R1 API to generate document-conditioned retrieval queries with task-specific prompts, detailed in Appendix~\ref{app:synthetic-data}.
We then reuse the Stage I teacher-scoring and score-stratified sampling pipeline to produce pseudo ranking supervision for the pairwise objective in Eq.~\eqref{eq:rank_loss}.

This stage provides label-free, teacher-distilled adaptation toward target ranking distributions without using human relevance annotations or evaluation labels from the target tasks.
Rather than training isolated task-specific rerankers and merging them afterward, the unified teacher-scoring pipeline converts task-specific synthetic queries into a shared multi-task ranking objective, allowing knowledge to be shared and transferred across diverse multilingual corpus distributions before checkpoint merging.

\subsection{Stage III: Model Merging}
After the first two stages, we obtain multiple checkpoints from different data mixtures and training runs. To produce a single deployable reranker, we merge selected checkpoints with spherical linear interpolation (SLERP). Given two corresponding parameter tensors $\theta_a$ and $\theta_b$, viewed as vectors, SLERP computes:
\begin{equation}
\begin{aligned}
    \mathrm{slerp}(\theta_a,\theta_b;t)
    &=
    \frac{\sin((1-t)\Omega)}{\sin \Omega}\theta_a
    +
    \frac{\sin(t\Omega)}{\sin \Omega}\theta_b, \\
    \Omega
    &=
    \arccos
    \left(
    \frac{\theta_a^{\top}\theta_b}
    {\|\theta_a\|\|\theta_b\|}
    \right),
\end{aligned}
\label{eq:slerp}
\end{equation}
where $t\in[0,1]$ is the interpolation coefficient. For an ordered sequence of checkpoints $\{\theta_i\}_{i=1}^{K}$ with weights $\{w_i\}_{i=1}^{K}$, we initialize $\theta^{(1)}=\theta_1$ and sequentially compute
\begin{equation}
\begin{aligned}
\theta^{(i)}
&=
\mathrm{slerp}\!\left(
\theta^{(i-1)},\theta_i;t_i
\right), \\
t_i
&=
\frac{w_i}{\sum_{j=1}^{i}w_j},
\qquad i=2,\ldots,K.
\end{aligned}
\label{eq:sequential-slerp}
\end{equation}
The final merged model is $\theta^{(K)}$, which requires no additional training and introduces no inference-time ensembling cost.

\section{Experiments}

\subsection{Experimental Setup}

\paragraph{Datasets.}
Our main evaluation covers 8 MMTEB reranking tasks from MTEB-multilingual-v2 and MTEB-eng-v2: six multilingual tasks (AlloprofReranking, RuBQReranking, T2Reranking, VoyageMMarcoReranking, WebLINXCandidatesReranking, and WikipediaRerankingMultilingual) and two English tasks (AskUbuntuDupQuestions and MindSmallReranking).
We further evaluate general English reranking on 13 BEIR tasks and multilingual reranking on MIRACL across 18 languages.

\paragraph{Evaluation Procedure.}
For MMTEB, we follow the official evaluation protocol and report each task's standard metric. For BEIR and MIRACL, we use Qwen3-Embedding-0.6B\footnote{\href{https://huggingface.co/Qwen/Qwen3-Embedding-0.6B}{Qwen3-Embedding-0.6B}} as a shared first-stage retriever, rerank the same top-100 candidates for controlled comparison, and report nDCG@10. Best and second-best scores are marked in bold and underlined, respectively.

\subsection{Main Results}
\begin{table*}[!t]
\centering
\caption{Reranking performance on the MMTEB benchmark.
Results are collected from the MTEB leaderboard%
\textsuperscript{\emph{c}}.}
\label{tab:mmteb_results}
\setlength{\tabcolsep}{2pt}
\small
\resizebox{\textwidth}{!}{%
\begin{tabular}{lrrrrrrrrrrr}
\toprule
\multirow{2}{*}{Model} & \multicolumn{7}{c}{\textit{MTEB-multilingual-v2}} & \multicolumn{3}{c}{\textit{MTEB-eng-v2}} & \multirow{2}{*}{Avg.} \\
\cmidrule(lr){2-8} \cmidrule(lr){9-11}
& Allo & RuBQ & T2 & Voyage & WebLINX & Wiki & Avg. & Ask & Mind & Avg. & \\
\midrule
\multicolumn{12}{c}{\emph{Small-Scale Models ($<4$B)}} \\
\midrule
F2LLM-v2-330M
& 75.91 & 69.63 & 66.64 & 69.48 & 11.18 & 87.25 & 63.35 & 63.00 & 30.99 & 47.00 & 59.26 \\
F2LLM-v2-0.6B
& 78.39 & 71.52 & 67.00 & 72.57 & 13.61 & 87.73 & 65.14 & 64.26 & 31.10 & 47.68 & 60.77 \\
jina-embeddings-v5-omni-small
& 81.39 & 74.89 & 68.04 & 68.84 & 11.33 & 89.46 & 65.66 & 66.08 & 32.69 & 49.38 & 61.59 \\
jina-embeddings-v5-text-small
& 81.39 & 74.89 & 68.04 & 68.84 & 11.33 & 89.46 & 65.66 & 66.08 & 32.69 & 49.38 & 61.59 \\
jina-reranker-v3
& 82.34 & 79.01 & \textbf{70.28} & 73.52 & 8.80 & \textbf{93.08} & 67.84 & 66.00 & -- & -- & -- \\
gte-Qwen2-1.5B-instruct
& 73.49 & 73.88 & 67.43 & 65.22 & 8.16 & 87.31 & 62.58 & 64.55 & \textbf{33.94} & 49.25 & 59.25 \\
F2LLM-v2-1.7B
& 80.11 & 75.33 & 66.96 & 75.94 & 15.62 & 89.08 & 67.17 & 65.98 & 32.49 & 49.24 & 62.69 \\
Giga-Embeddings-A1.8B\textsuperscript{\emph{a}}
& 81.06 & \textbf{80.88} & 66.98 & 71.20 & 14.53 & \underline{92.09}
& 67.79 & 67.68 & 30.94 & 49.31 & 63.17 \\

Giga-Embeddings-3B\textsuperscript{\emph{b}}
& 80.15 & 79.88 & 66.64 & 70.60 & 14.72 & 91.59
& 67.26 & 68.53 & 30.72 & 49.62 & 62.85 \\
\midrule
\textbf{Querit-Reranker-A0.4B}
& 79.39 & 75.37 & 68.98 & 68.59 & 11.63 & 91.05 & 65.84 & 64.02 & 32.52 & 48.27 & 61.44 \\
\midrule
\multicolumn{12}{c}{\emph{Medium-Scale Models (4B--8B)}} \\
\midrule
Qwen3-Embedding-4B
& \underline{85.13} & 72.28 & 67.27 & 65.61 & 11.30 & 88.89 & 65.08 & \underline{68.81} & 32.71 & \underline{50.76} & 61.50 \\
F2LLM-v2-4B
& 82.25 & 76.43 & 67.27 & \underline{81.74} & 19.19 & 89.40 & 69.38 & 67.12 & 33.01 & 50.07 & 64.55 \\
gte-Qwen2-7B-instruct
& 81.10 & 74.13 & 67.80 & 70.06 & 12.50 & 87.73 & 65.55 & 67.58 & 33.36 & 50.47 & 61.78 \\
Qwen3-Embedding-8B
& 84.88 & 73.47 & 68.29 & 65.08 & 11.79 & 90.28 & 65.63 & \textbf{70.20} & 32.93 & \textbf{51.56} & 62.12 \\
llama-embed-nemotron-8b
& 81.29 & \underline{80.49} & 68.29 & 71.25 & 13.66 & 91.68 & 67.78 & -- & -- & -- & -- \\
Octen-Embedding-8B
& \textbf{85.40} & 76.88 & 67.14 & 68.00 & 17.41 & 91.00 & 67.64 & -- & -- & -- & -- \\
F2LLM-v2-8B
& 82.71 & 77.61 & 67.41 & \textbf{82.17} & \underline{22.46} & 89.65 & \underline{70.34} & 66.99 & \underline{33.43} & 50.21 & \underline{65.30} \\
\midrule
\textbf{Querit-Reranker-4B}
& 83.89 & 79.84 & \underline{69.39} & 70.74 & \textbf{30.81} & 91.78 & \textbf{71.08} & 66.25 & 32.15 & 49.20 & \textbf{65.61} \\
\bottomrule
\end{tabular}
}
\vspace{2pt}
\parbox{\textwidth}{%
\footnotesize
\textsuperscript{\emph{a}}
Giga-Embeddings-instruct-10B-A1.8B-0826;
\quad
\textsuperscript{\emph{b}}
Giga-Embeddings-instruct-3B-0826.
\par
\textsuperscript{\emph{c}}
\href{https://huggingface.co/spaces/mteb/leaderboard}
{MTEB Multilingual leaderboard} snapshot accessed on August 29, 2026.
}
\end{table*}

\begin{figure}[t]
\centering
\includegraphics[width=\linewidth]{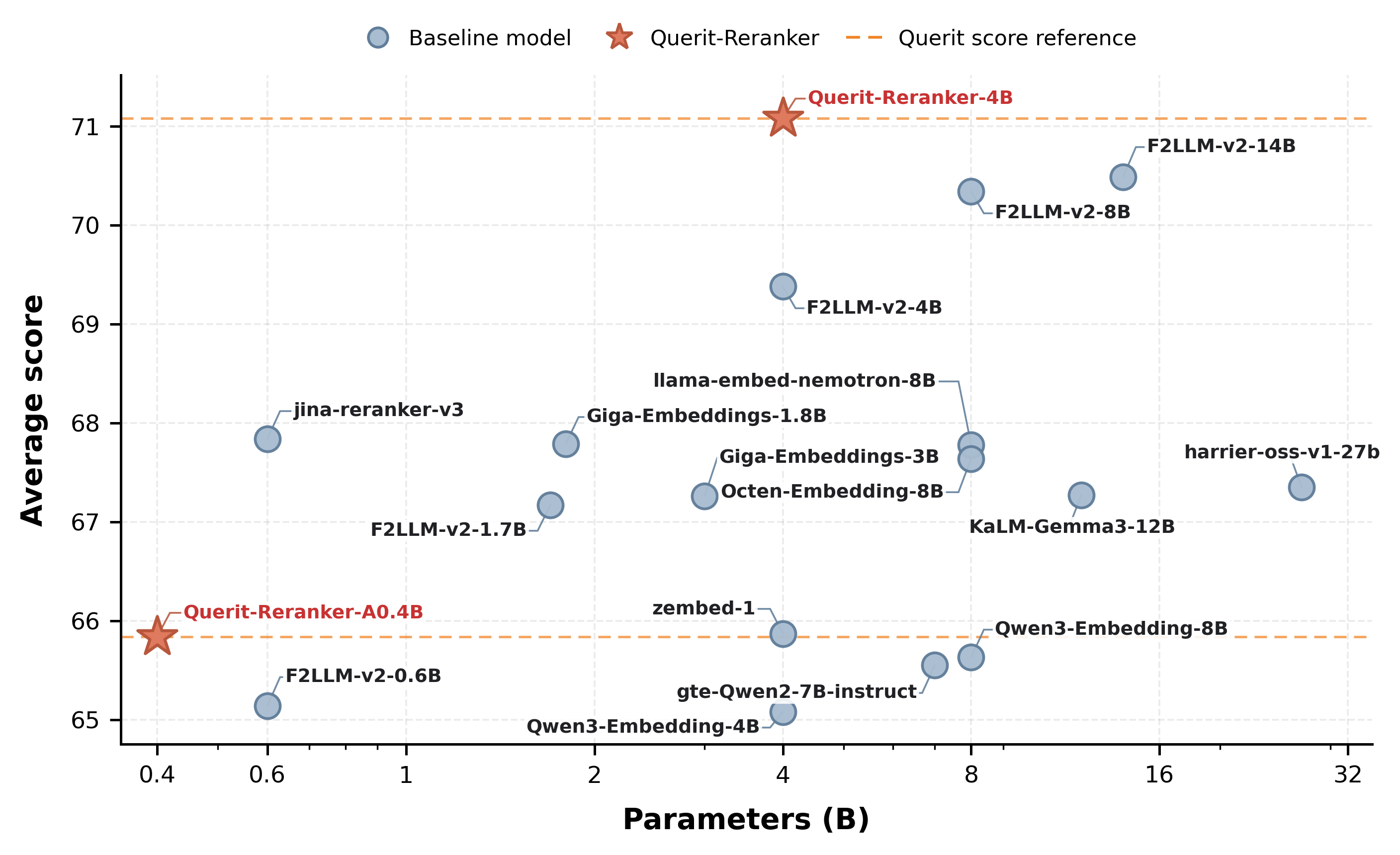}
\caption{Performance--parameter trade-offs on MTEB-multilingual-v2 reranking tasks.}
\label{fig:param_tradeoff}
\end{figure}

Table~\ref{tab:mmteb_results} reports the main results on the 8 reranking tasks from MMTEB. We group representative models with available leaderboard results by parameter scale, focusing on models up to 8B. 

Querit-Reranker-A0.4B shows strong parameter efficiency among small-scale models. Despite its compact active-parameter scale, it achieves an 8-task average of 61.44, outperforming F2LLM-v2-0.6B (60.77) and gte-Qwen2-1.5B-instruct (59.25), and reaches 65.84 on MTEB-multilingual-v2. This suggests that the proposed training recipe can equip compact Querit-based rerankers with competitive multilingual reranking capability.

Scaling the same recipe to the 4B backbone further improves performance.
Querit-Reranker-4B achieves the best average among the reported models on both MTEB-multilingual-v2 (71.08) and the combined 8-task setting (65.61), surpassing larger 8B-scale systems, while also obtaining the highest WebLINXCandidatesReranking score of 30.81.
These results indicate that the proposed recipe scales effectively from the compact Querit-based reranker to the 4B backbone.

Figure~\ref{fig:param_tradeoff} further visualizes the performance--parameter trade-off on MTEB-multilingual-v2. Beyond the up-to-8B models in Table~\ref{tab:mmteb_results}, Querit-Reranker-4B also outperforms larger 14B--27B systems, such as F2LLM-v2-14B and harrier-oss-v1-27b, achieving a \textbf{SOTA} average score of 71.08 on the official MTEB-Multilingual-v2 Reranking Tasks Leaderboard as of August 29, 2026. The full leaderboard snapshots corresponding to Figure~\ref{fig:param_tradeoff} and the complete results are provided in Table~\ref{tab:mteb_multi_reranking_results} and Table~\ref{tab:reranking_mteb_merged_mmteb_eng_complete} in Appendix~\ref{app:detailed-results}.

\subsection{Stage-wise and Merging Ablations}
\label{sec:stage-merging-ablation}

\paragraph{Merging strategy comparison.} Our task-adapted checkpoints share the same architecture and initialization but capture complementary capabilities from different data mixtures. We select the best-performing task-adapted checkpoint for each of Allo, RuBQ, T2, Voyage, WebLINX, and Wiki; checkpoint weights reported below follow this order. We compare five training-free strategies that produce a single model without runtime ensembling. Linear interpolation merges all parameters with weights $1{:}2{:}1{:}1{:}1{:}1$. Scoring-head replacement combines the general reranker backbone with the WebLINX-oriented scoring head, while top-layer interpolation linearly merges their top four transformer layers and scoring heads using $\alpha=0.2$, retaining the lower layers of the general reranker. TIES-Merging constructs task vectors relative to the shared initialization, retains the top-$20\%$ largest-magnitude updates, resolves sign conflicts, and merges the aligned updates with $\lambda=1.0$. SLERP sequentially merges the six checkpoints with weights $1{:}3{:}1{:}1{:}1{:}1$ using Eqs.~\eqref{eq:slerp}--\eqref{eq:sequential-slerp}. As shown in Table~\ref{tab:merging-comparison}, SLERP achieves the best corresponding averages of 65.84, 48.27, and 61.44 while adding no training or inference overhead. These results motivated its use for Querit-Reranker-4B and kept the deployment pipeline consistent across model scales. Complete task-level results are reported in Table~\ref{tab:merging-comparison-full}.
\begin{table}[t]
\centering
\small
\setlength{\tabcolsep}{2pt}
\renewcommand{\arraystretch}{1.08}
\caption{Merging-method ablation for Querit-Reranker-A0.4B.}
\label{tab:merging-comparison}
\begin{tabular*}{\columnwidth}{
@{\extracolsep{\fill}}
lrrr
}
\toprule
\textbf{Merging strategy}
& \textbf{Multi.}
& \textbf{Eng.}
& \textbf{Overall} \\
\midrule

Linear interpolation
& 65.58 & 47.02 & 60.94 \\

Scoring-head replacement
& 65.12 & 46.19 & 60.38 \\

Top-4 layers + head
& 65.56 & 46.62 & 60.82 \\

TIES
& 65.75 & 47.65 & 61.23 \\

\textbf{SLERP}
& \textbf{65.84}
& \textbf{48.27}
& \textbf{61.44} \\

\bottomrule
\end{tabular*}
\end{table}

\paragraph{Stage-wise ablation.} Table~\ref{tab:stage-wise-ablation} reports the aggregate contribution of each training stage at both model scales. For A0.4B, the overall average increases from 26.56 to 56.98 after Stage I and further improves to 60.90 and 61.44 after Stages II and III, respectively. Stage I accounts for most of the improvement because the general-purpose backbone must first acquire broad query--document relevance modeling through multilingual teacher supervision. For 4B, whose initialization is already retrieval-oriented, the corresponding gains are more evenly distributed across stages, with the overall average increasing from 61.50 to 63.13, 64.07, and 65.61. The multilingual average improves consistently across all stages for both model scales. Although the English average of 4B does not improve monotonically, the gains on the multilingual benchmark lead to consistent improvements in its overall average. Complete task-level results are reported in Table~\ref{tab:stage-wise-ablation-full}.
\begin{table}[t]
\centering
\caption{Stage-wise ablation of Querit-Reranker. For Stage II, we report the strongest individual pre-merging checkpoint according to the eight-task average.}
\label{tab:stage-wise-ablation}
\setlength{\tabcolsep}{2pt}
\small
\resizebox{\columnwidth}{!}{%
\begin{tabular}{llrrr}
\toprule
\textbf{Model}
& \textbf{Stage}
& \textbf{Multi. Avg.}
& \textbf{Eng. Avg.}
& \textbf{Overall Avg.} \\
\midrule

\multirow{4}{*}{\textbf{A0.4B}}
& Base
& 25.03
& 31.13
& 26.56 \\

& Stage I
& 61.16
& 44.41
& 56.98 \\

& Stage II
& 65.31
& 47.66
& 60.90 \\

& \textbf{Stage III}
& \textbf{65.84}
& \textbf{48.27}
& \textbf{61.44} \\

\midrule

\multirow{4}{*}{\textbf{4B}}
& Base
& 65.08
& 50.76
& 61.50 \\

& Stage I
& 67.78
& 49.19
& 63.13 \\

& Stage II
& 69.20
& 48.66
& 64.07 \\

& \textbf{Stage III}
& \textbf{71.08}
& \textbf{49.20}
& \textbf{65.61} \\

\bottomrule
\end{tabular}%
}
\end{table}

\begin{table*}[h]
\centering
\caption{Reranking performance on the BEIR benchmark measured by nDCG@10.}
\label{tab:beir_results}
\resizebox{\textwidth}{!}{%
\begin{tabular}{lcccccccccccccc}
\toprule
Models & Avg. & TC & NFC & NQ & HQA & FQA & AA & TCH & DBP & SD & FVR & CFV & SF & QRA \\
\midrule
\multicolumn{15}{c}{\emph{First-stage Retriever}} \\
\midrule
Qwen3-Embedding-0.6B 
& 54.11 & 88.64 & 36.34 & 52.62 & 64.59 & 45.08 & 47.90 & 28.73 & 39.06 & 22.90 & 85.82 & 35.23 & 69.67 & 86.86 \\

\midrule
\multicolumn{15}{c}{\emph{Second-stage Reranker}} \\
\midrule
Qwen3-Reranker-0.6B 
& 48.09 & 80.15 & 36.40 & 35.71 & 73.28 & 30.98 & 40.66 & 21.63 & 33.42 & 17.58 & 79.62 & 23.67 & 72.05 & 80.00 \\

bge-reranker-v2-m3
& 56.42 & 84.74 & 34.51 & 68.40 & 77.56 & 44.34 & 33.84 & \underline{35.10} & 47.05 & 18.20 & 89.43 & 37.42 & 73.86 & \underline{89.19} \\

jina-reranker-v3
& 61.71 & 87.06 & 38.70 & \underline{71.84} & \underline{78.10} & 49.51 & \textbf{72.85} & 33.70 & 47.77 & 23.14 & 93.20 & 38.95 & 77.00 & \textbf{90.38} \\

Qwen3-Reranker-4B
& 59.92 & 88.75 & 41.35 & 63.18 & 76.55 & 50.71 & 52.45 & \textbf{35.80} & 50.49 & \underline{25.46} & 86.33 & \underline{44.82} & 78.60 & 84.53 \\

Qwen3-Reranker-8B
& \underline{61.81} & \underline{89.66} & \underline{42.79} & 67.20 & 78.05 & \textbf{57.51} & \underline{56.80} & 33.66 & \underline{50.84} & \textbf{27.49} & 87.65 & \textbf{46.37} & 78.24 & 87.33 \\

\midrule
Querit-Reranker-A0.4B
& 59.28 & 88.72 & 41.41 & 66.56 & 76.31 & 48.05 & 46.61 & 33.82 & 47.85 & 22.84 & \underline{93.70} & 37.29 & \underline{78.87} & 88.63 \\
Querit-Reranker-4B
& \textbf{62.29} & \textbf{89.82} & \textbf{43.35} & \textbf{74.21} & \textbf{79.59} & \underline{56.16} & 55.89 & 31.42 & \textbf{52.80} & 23.64 & \textbf{94.45} & 39.04 & \textbf{81.22} & 88.18 \\

\bottomrule 
\end{tabular} 
} 
\end{table*}
\begin{table*}[!t]
\centering
\caption{Reranking performance on the MIRACL benchmark measured by nDCG@10.}
\label{tab:miracl_results}
\setlength{\tabcolsep}{2pt}
\small
\resizebox{\textwidth}{!}{%
\begin{tabular}{lrrrrrrrrrrrrrrrrrrr}
\toprule
Models & Avg. & AR & BN & EN & ES & FA & FI & FR & HI & ID & JA & KO & RU & SW & TE & TH & ZH & DE & YO \\
\midrule
\multicolumn{20}{c}{\emph{First-stage Retriever}} \\
\midrule
Qwen3-Embedding-0.6B
& 59.87 & 68.81 & 66.20 & 50.64 & 51.93 & 53.33 & 68.92 & 52.81 & 51.28 & 51.71 & 62.99 & 60.36 & 58.90 & 46.80 & 77.21 & 73.63 & 58.03 & 51.58 & 72.59 \\
\midrule
\multicolumn{20}{c}{\emph{Second-stage Reranker}} \\
\midrule
Qwen3-Reranker-0.6B
& 59.59 & 65.80 & 66.88 & 50.33 & 49.68 & 53.96 & 68.43 & 49.87 & 52.99 & 49.94 & 62.80 & 62.72 & 59.36 & 51.45 & 74.94 & 74.52 & 55.93 & 50.95 & 72.11 \\

bge-reranker-v2-m3
& 67.64 & 75.51 & 77.36 & 57.81 & \underline{56.45} & 60.35 & 77.39 & \underline{59.64} & 58.80 & \textbf{56.37} & 71.88 & 66.01 & 68.06 & 61.11 & \textbf{85.88} & 81.74 & 62.39 & 57.55 & 83.19 \\

jina-reranker-v3
& 66.83 & 75.23 & 75.91 & 57.80 & 55.83 & 59.14 & 76.61 & 57.93 & 56.61 & \underline{56.26} & 71.84 & 67.27 & 67.35 & 59.72 & \underline{84.15} & 81.85 & 64.23 & 56.64 & 78.51 \\

Qwen3-Reranker-4B
& 67.47 & 74.99 & 76.82 & 58.78 & 54.33 & 62.22 & \underline{78.87} & 57.96 & 61.64 & 53.78 & 70.73 & 67.25 & 68.90 & 63.83 & 83.50 & 78.85 & 61.86 & 59.94 & 80.21 \\

Qwen3-Reranker-8B
& \underline{68.97} & 77.33 & \underline{78.99} & 58.25 & 55.93 & \underline{62.87} & 77.26 & 57.87 & \textbf{63.92} & 55.14 & 72.52 & 69.13 & 69.78 & \underline{68.08} & 83.34 & \underline{82.49} & 63.86 & \underline{60.39} & \textbf{84.36} \\

\midrule
\textbf{Querit-Reranker-A0.4B}
& 67.70 & \underline{77.65} & 78.53 & \underline{61.62} & 55.38 & 60.17 & 75.61 & 58.60 & 62.11 & 54.02 & \underline{74.22} & \underline{69.86} & \underline{71.69} & 63.01 & 74.76 & 77.06 & \underline{66.69} & 59.57 & 77.98 \\

\textbf{Querit-Reranker-4B}
& \textbf{71.13} & \textbf{79.34} & \textbf{81.28} & \textbf{62.20} & \textbf{58.41} & \textbf{62.98} & \textbf{80.94} & \textbf{60.45} & \underline{63.71} & 55.94 & \textbf{76.96} & \textbf{70.52} & \textbf{74.47} & \textbf{72.70} & 83.44 & \textbf{82.98} & \textbf{67.52} & \textbf{62.18} & \underline{84.31} \\
\bottomrule
\end{tabular}
}
\end{table*}

\subsection{General English Reranking Performance on BEIR}
Table~\ref{tab:beir_results} reports general English reranking results on BEIR measured by nDCG@10.
Querit-Reranker-A0.4B achieves an average score of 59.28, improving the first-stage retriever by 5.17 points and clearly outperforming Qwen3-Reranker-0.6B and bge-reranker-v2-m3. Despite its compact active-parameter scale, it remains competitive with larger rerankers and obtains second-best results on FEVER and SciFact.

Querit-Reranker-4B further improves the average to 62.29, achieving the best overall performance among all compared models. It outperforms strong baselines such as jina-reranker-v3 and Qwen3-Reranker-4B, surpasses the larger Qwen3-Reranker-8B by 0.48 points, and ranks first on 7 of 13 BEIR tasks.
Overall, these results demonstrate effective relevance refinement across diverse English retrieval tasks and show that the training recipe scales from the compact Querit-based reranker to the 4B backbone.

\subsection{Multilingual Performance on MIRACL}
Table~\ref{tab:miracl_results} reports multilingual reranking performance on MIRACL.
Using Qwen3-Embedding-0.6B as the shared first-stage retriever, Querit-Reranker-A0.4B improves the average nDCG@10 from 59.87 to 67.70, while Querit-Reranker-4B further raises it to 71.13.
These gains show that Querit-Reranker provides substantial second-stage refinement for multilingual retrieval.

Querit-Reranker-4B achieves the best overall performance among all compared rerankers, outperforming Qwen3-Reranker-8B by 2.16 points on average despite using a smaller 4B backbone.
Meanwhile, Querit-Reranker-A0.4B achieves the second-best result on 6 of 18 languages and outperforms other 0.6B-scale rerankers.
Overall, MIRACL results confirm the effectiveness and scalability of Querit-Reranker for multilingual reranking.

\section{Conclusion}
\label{sec:conclusion}

We presented \textbf{Querit-Reranker}, a family of multilingual cross-encoder rerankers for standard retrieve-then-rerank pipelines. Querit-Reranker uses a staged reranking-oriented training recipe to build broad multilingual relevance matching ability, perform label-free multi-task adaptation toward target ranking distributions, and consolidate complementary checkpoints with SLERP. Experiments on MMTEB, BEIR, and MIRACL show that this recipe is effective for both the compact Querit-Reranker-A0.4B and the larger Querit-Reranker-4B, providing a practical path toward efficient and strong multilingual reranking.

\clearpage

\section{Ethical Considerations}

\paragraph{Data and privacy.}
Our training data include publicly available ranking datasets and de-identified private search data from real-world business scenarios. The private data were de-identified before model development and handled under the data-governance and access-control procedures of the data provider; no direct personal identifiers were intentionally used for training. Public datasets, pretrained models, and teacher models are used subject to their respective licenses and terms of use.

\paragraph{Bias and downstream use.}
Teacher-provided scores and LLM-generated queries may inherit biases from the underlying models and source corpora, including uneven representation across languages, regions, and domains. Although the deployed reranker does not directly generate text, it can affect information exposure by promoting or suppressing retrieved content and may amplify biases present in the candidate corpus or first-stage retriever. Its relevance scores should not be interpreted as indicators of factuality, safety, or fairness. We recommend task-specific evaluation and bias auditing before deployment, particularly in high-stakes or user-facing applications.

\paragraph{Model release and resource use.}
We release the model weights for research purposes. Users are responsible for complying with the licenses of the models, datasets, and downstream content involved in their applications. Although the compact architecture and training-free checkpoint merging avoid runtime ensembling overhead, model training and teacher inference incur computational costs.
\section{Limitations}
Although Querit-Reranker demonstrates strong multilingual reranking performance, it does not explicitly optimize a dedicated cross-lingual reranking or alignment objective.
Its multilingual capability mainly comes from Stage I, which combines open-source multilingual ranking data with de-identified private search data from real-world business scenarios, and Stage II, which further performs task-targeted data mining on representative multilingual reranking corpora.
However, our training data do not include explicitly constructed cross-lingual reranking pairs, and our training objective does not incorporate explicit language-alignment losses.
As a result, the model may still exhibit uneven behavior across low-resource languages or cross-lingual retrieval scenarios.
In future work, constructing cross-lingual reranking data and introducing explicit alignment-oriented objectives may further improve the model's multilingual understanding and robustness across languages.

In addition, due to resource constraints, our high-quality synthetic-data mining stage only covers four representative MTEB-multilingual-v2 reranking corpora: AlloprofReranking, RuBQReranking, T2Reranking, and VoyageMMarcoReranking. These datasets expose the model to French, Russian, Chinese, and Japanese reranking distributions, but they do not cover all languages and reranking tasks in MMTEB. Therefore, the effect of extending synthetic-query mining to the full set of MMTEB reranking tasks remains to be explored. Future work will investigate broader multilingual synthetic data construction and explicit cross-lingual alignment to further improve robustness across languages and retrieval settings.

Regarding evaluation, the WebLINX results in this comparison reflect the historical MTEB leaderboard setting and the official evaluation implementation available at that time. A subsequently identified MRR tie-breaking issue in the evaluator may affect historical WebLINX results. The reported values should therefore be interpreted in the context of this historical setting.

\section{Acknowledgements}
This work was funded by the National Natural Science Foundation of China (NSFC) under Grant Nos. 62372431, 62472408, U25B2076, and 62441229; the National Key Research and Development Program of China under Grant No. 2025ZD0123301; and the Strategic Priority Research Program of the CAS under Grant Nos. XDB0680301 and XDB0680102.

\bibliography{references}

% \bibliography{custom}

\clearpage
\appendix
\makeatletter
\setlength{\@dblfptop}{0pt}
\makeatother
\section{Targeted Synthetic Data Construction}
\label{app:synthetic-data}

This appendix provides the detailed prompts used for high-quality synthetic data mining.
We construct synthetic queries for four representative MTEB-multilingual-v2 reranking tasks: AlloprofReranking, RuBQReranking, T2Reranking, and VoyageMMarcoReranking.
These tasks cover French, Russian, Chinese, and Japanese, respectively, and differ substantially in domain, document style, and query intent.
Therefore, we design task-specific prompts rather than using a single generic query-generation template.
For each task, the generated queries are parsed with special query boundary tokens and then used in the teacher-scoring and score-stratified sampling pipeline described in Section~\ref{sec:model-training}.

\subsection{AlloprofReranking}

AlloprofReranking is a French educational reranking task based on K12-style learning content.
The prompt is designed to generate informal and conversational French queries that resemble questions students may ask in an online forum.

\begin{promptbox}
\begin{Verbatim}[fontsize=\small,breaklines=true,breakanywhere=true]
You are an information retrieval assistant for a French K12 educational platform.

Your task is to generate exactly FIVE user search queries that K12 students might naturally ask on an online forum.
The query should be written in informal, conversational French, similar to real student questions.

The generated query should:
1. Be a natural and realistic question that a K12 student could plausibly ask.
2. Reflect the main information needed from the given document, without copying or restating its text.
3. Use informal, everyday French rather than formal or academic language.
4. Be concise, between 10 and 100 words.

Output requirements:
- Output exactly FIVE queries.
- Wrap the query between the special tokens <|begin_of_query|> and <|end_of_query|>.
- Do not output anything outside these tokens.

Document:
{content}
\end{Verbatim}
\end{promptbox}

\subsection{RuBQReranking}

RuBQReranking is a Russian factoid-style reranking task based on Wikipedia-like documents.
Since many documents focus on named entities, the prompt encourages factual questions and asks the generator to use alternative entity surface forms when available, making the generated queries closer to realistic retrieval behavior.

\begin{promptbox}
\begin{Verbatim}[fontsize=\small,breaklines=true,breakanywhere=true]
You are an information retrieval assistant generating search queries for a Russian Wikipedia-style factoid QA dataset.

You will be given:
- A Russian document in a Wikipedia-like style.
- The document typically focuses on a single main entity (person, organization, city, work, event).

Your task:
Generate exactly ONE natural Russian query that asks for ONE factual piece of information
that is directly answered in the document (typically within the first 1-3 sentences).

CRITICAL REQUIREMENTS (must follow ALL when applicable):
1) Identify the main named entity the document is about.
2) If there exists ANY plausible alternative surface form for that entity across writing systems or common usage
   (Latin vs Cyrillic, phonetic transliteration, common colloquial form, mixed form),
   then the query MUST use a surface form DIFFERENT from the one appearing in the document.
3) Do NOT use the exact entity surface form from the document when a reasonable alternative exists.
4) If no alternative exists, use the canonical Russian form.

Query constraints:
- Russian only.
- One entity, one fact, one question.
- Prefer: «Когда...», «Кто...», «В каком году...», «Где...».
- 5-20 words.
- Do NOT copy or paraphrase any sentence from the document.

STRICT OUTPUT RULES (mandatory):
- Output EXACTLY ONE LINE.
- The line must be exactly in this format:
  <|begin_of_query|> ... <|end_of_query|>
- Do NOT output any explanations, rationales, comments, HTML, markdown, or extra text.
- Do NOT output multiple candidates.

Document:
{content}
\end{Verbatim}
\end{promptbox}

\subsection{T2Reranking}

T2Reranking contains Chinese documents with diverse topics and noisy web-style formatting.
The prompt, therefore, emphasizes core answerable information, robustness to HTML and navigation noise, and realistic short Chinese search queries.

\begin{promptbox}
\begin{Verbatim}[fontsize=\small,breaklines=true,breakanywhere=true]
你是一名信息检索助手，需要为中文文档生成搜索查询。

文档特点说明：
- 文档为中文，主题多为生活常识、学习问题、社会议题等。
- 文本形式口语化，常为短句 + 若干关键词拼接。
- 文档中可能包含大量 HTML 标签（如 <br>、<table>、<h2>、<img> 等）、导航标题、列表、栏目堆叠、广告信息等噪声。
- 有些文档是结构化信息（如时间表、号码列表、步骤教程等）。

你的任务是：
基于文档的“核心可回答信息”，生成三个真实用户可能输入搜索引擎的中文查询。

生成的查询必须满足：

1. 每个查询只围绕一个具体、明确的信息点（例如：时间、价格、号码、定义、步骤、原因、人物、关系等）。
2. 查询应模拟真实搜索行为，简洁、直接、口语化，而不是完整句子式提问。
3. 不要被 HTML 标签、栏目标题堆叠、导航词（如“更多”“相关推荐”“相关工具”等）误导。
4. 如果文档中包含明确的数字信息（如电话、价格、分享码、时间、日期、数量等），至少有一个查询必须直接询问该数字。
5. 查询应紧贴文档核心实体或主题关键词（允许词面重合）。
6. 避免生成过于宽泛或总结性的问题（例如“这篇文章讲了什么”）。

输出要求：
- 输出三条查询。
- 每条查询必须严格包裹在 <|begin_of_query|> 和 <|end_of_query|> 之间。
- 不要输出任何额外说明或解释文本。

Document:
{content}
\end{Verbatim}
\end{promptbox}

\subsection{VoyageMMarcoReranking}

VoyageMMarcoReranking is a Japanese MS MARCO-style reranking task.
The prompt asks the generator to produce concise and realistic Japanese search queries, with special attention to concrete information needs such as definitions, attributes, facts, and numerical values.

\begin{promptbox}
\begin{Verbatim}[fontsize=\small,breaklines=true,breakanywhere=true]
You are an information retrieval assistant generating queries for a Japanese MS MARCO-style dataset.

Your task is to generate exactly FIVE user search queries that a typical user might naturally type into a search engine.
The queries should be written in natural Japanese and reflect realistic search behavior.

The generated queries should:
1. Ask about one specific and concrete piece of information contained in the document.
2. Reflect a single, clear information need (e.g., definition, fact, number, attribute, cause, person, etc.).
3. Be concise and focused, typically short and direct rather than open-ended.
4. Closely align with the key entity or concept in the document (lexical overlap is acceptable).
5. If the document contains an important number (e.g., phone number, routing number, price, temperature, quantity, date), at least ONE query should directly ask for that number.

Output requirements:
- Output exactly FIVE queries.
- Each query must be wrapped between the special tokens <|begin_of_query|> and <|end_of_query|>.
- Do not output anything outside these tokens.
- Do not add explanations or additional text.

Document:
{content}
\end{Verbatim}
\end{promptbox}

\section{Training Data Statistics}
\label{app:training-data-statistics}

Table~\ref{tab:synthetic-data-statistics} summarizes the data used in the two data-mining stages.

\begin{table}[H]
\centering
\footnotesize
\setlength{\tabcolsep}{3pt}
\renewcommand{\arraystretch}{1.18}
\caption{Statistics of training data used in each stage.}
\label{tab:synthetic-data-statistics}
\begin{tabularx}{\linewidth}{
    >{\raggedright\arraybackslash}p{0.25\linewidth}
    >{\raggedright\arraybackslash}X
    >{\raggedright\arraybackslash}p{0.20\linewidth}
}
\toprule
\textbf{Stage} & \textbf{Dataset} & \textbf{Size} \\
\midrule
\textbf{Stage I}\newline High-quality data collection
&
MSMARCO, MLDR, Mr.TyDi, T2Ranking, CQADupStack, MIND-small, Ruri, MIRACL, and private high-quality data
&
Open-source: $\sim$3.52M\newline
Private: $\sim$2.05M
\\
\midrule
\textbf{Stage II}\newline Targeted synthetic data mining
&
Targeted synthetic data for AlloprofReranking, VoyageMMarcoReranking, T2Reranking, and RuBQReranking
&
$\sim$9.4M
\\
\bottomrule
\end{tabularx}
\end{table}

Stage I collects broad reranking supervision from open-source retrieval datasets and private high-quality search data collected from real-world business scenarios. The open-source portion includes MS MARCO \cite{nguyen2016msmarco}, MLDR \cite{chen2024bgem3}, Mr.TyDi \cite{mrtydi,tydiqa}, T2Ranking \cite{xie2023t2ranking}, CQADupStack \cite{hoogeveen2015cqadupstack}, MIND-small \cite{wu-etal-2020-mind}, Ruri \cite{tsukagoshi2024rurijapanesegeneraltext}, and MIRACL \cite{zhang2023miracl}. For all open-source datasets, we use only their official training splits and exclude validation and test splits from training. Following standard industrial data-curation practices, the private data were de-identified before use, with personally identifiable information such as names and phone numbers removed. The resulting private dataset does not contain direct user identifiers and covers diverse retrieval settings such as web search, multilingual retrieval, question answering, duplicate question detection, and news recommendation.

Stage II further constructs targeted synthetic data for four representative MTEB-multilingual-v2 reranking tasks, namely AlloprofReranking, VoyageMMarcoReranking, T2Reranking, and RuBQReranking. This targeted stage is designed to better match the corpus distributions and query styles of the multilingual reranking benchmark.

\begin{table*}[!t]
\centering
\caption{Top-20 reranking performance on the MTEB-multilingual-v2 benchmark based on the MTEB leaderboard snapshot accessed on August 29, 2026. Results are measured using the official task metrics, and the average is computed over all six reranking tasks. Best scores are in bold.}
\label{tab:mteb_multi_reranking_results}
\setlength{\tabcolsep}{4.5pt}
\small
\begin{tabular}{lcrrrrrrr}
\toprule
Model & Size & Allo & RuBQ & T2 & Voyage & WebLINX & Wiki & Avg. \\
\midrule
\textbf{Querit/Querit-Reranker-4B} & 4B & 83.89 & 79.84 & 69.39 & 70.74 & \textbf{30.81} & 91.78 & \textbf{71.08} \\
codefuse-ai/F2LLM-v2-14B & 14B & 83.44 & 78.30 & 67.22 & \textbf{83.66} & 19.99 & 90.30 & 70.48 \\
codefuse-ai/F2LLM-v2-8B & 8B & 82.71 & 77.61 & 67.41 & 82.17 & 22.46 & 89.65 & 70.34 \\
codefuse-ai/F2LLM-v2-4B & 4B & 82.25 & 76.43 & 67.27 & 81.74 & 19.19 & 89.40 & 69.38 \\
jinaai/jina-reranker-v3 & 0.6B & 82.34 & 79.01 & \textbf{70.28} & 73.52 & 8.80 & \textbf{93.08} & 67.84 \\
ai-sage/Giga-Embeddings-instruct-10B-A1.8B-0826 & 1.8B & 81.06 & \textbf{80.88} & 66.98 & 71.20 & 14.53 & 92.09 & 67.79 \\
nvidia/llama-embed-nemotron-8b & 8B & 81.29 & 80.49 & 68.29 & 71.25 & 13.66 & 91.68 & 67.78 \\
Octen/Octen-Embedding-8B & 8B & \textbf{85.40} & 76.88 & 67.14 & 68.00 & 17.41 & 91.00 & 67.64 \\
microsoft/harrier-oss-v1-27b & 27B & 83.18 & 77.72 & 66.65 & 67.24 & 18.85 & 90.43 & 67.35 \\
tencent/KaLM-Embedding-Gemma3-12B-2511 & 12B & 81.54 & 77.27 & 66.25 & 71.36 & 16.80 & 90.42 & 67.27 \\
ai-sage/Giga-Embeddings-instruct-3B-0826 & 3B & 80.15 & 79.88 & 66.64 & 70.60 & 14.72 & 91.59 & 67.26 \\
codefuse-ai/F2LLM-v2-1.7B & 1.7B & 80.11 & 75.33 & 66.96 & 75.94 & 15.62 & 89.08 & 67.17 \\
Bytedance/Seed1.6-embedding-1215 & N/A & 79.96 & 76.51 & 66.06 & 66.13 & 17.92 & 90.85 & 66.24 \\
zeroentropy/zembed-1 & 4B & 83.49 & 74.76 & 68.83 & 66.23 & 12.25 & 89.66 & 65.87 \\
\textbf{Querit/Querit-Reranker-A0.4B} & 0.4B & 79.39 & 75.37 & 68.98 & 68.59 & 11.63 & 91.05 & 65.84 \\
jinaai/jina-embeddings-v5-text-small & 0.6B & 81.39 & 74.89 & 68.04 & 68.84 & 11.33 & 89.46 & 65.66 \\
jinaai/jina-embeddings-v5-omni-small & 2B & 81.39 & 74.89 & 68.04 & 68.84 & 11.33 & 89.46 & 65.66 \\
Qwen/Qwen3-Embedding-8B & 8B & 84.88 & 73.47 & 68.29 & 65.08 & 11.79 & 90.28 & 65.63 \\
google/gemini-embedding-001 & N/A & 81.77 & 73.84 & 67.95 & 66.73 & 10.97 & 92.24 & 65.58 \\
Alibaba-NLP/gte-Qwen2-7B-instruct & 7B & 81.10 & 74.13 & 67.80 & 70.06 & 12.50 & 87.73 & 65.55 \\
% codefuse-ai/F2LLM-v2-0.6B & 0.6B & 78.39 & 71.52 & 67.00 & 72.57 & 13.61 & 87.73 & 65.14 \\
% Qwen/Qwen3-Embedding-4B & 4B & 85.13 & 72.28 & 67.27 & 65.61 & 11.30 & 88.89 & 65.08 \\
\bottomrule
\end{tabular}
\end{table*}
\section{Detailed Results}
\label{app:detailed-results}

This section provides task-level results supporting the performance--parameter trade-off analysis in Figure~\ref{fig:param_tradeoff}, the stage-wise ablation in Table~\ref{tab:stage-wise-ablation}, and the merging-method comparison in Table~\ref{tab:merging-comparison}.

\paragraph{Benchmark comparisons.} Table~\ref{tab:mteb_multi_reranking_results} reports the top-20 results from the six-task MTEB-multilingual-v2 reranking leaderboard snapshot, while Table~\ref{tab:reranking_mteb_merged_mmteb_eng_complete} reports the top-20 complete results in the eight-task MMTEB evaluation setting, including only models with available scores on all tasks. 

As shown in Table~\ref{tab:mteb_multi_reranking_results}, Querit-Reranker-4B achieves the best average score on the MTEB-multilingual-v2 reranking subset, reaching 71.08. It also obtains the highest score on WebLINXCandidatesReranking, with a score of 30.81, showing a clear advantage on this challenging web-oriented reranking task. Querit-Reranker-A0.4B achieves an average score of 65.84, outperforming several substantially larger embedding baselines and demonstrating strong parameter efficiency among compact models.
\begin{table*}[!t]
\centering
\caption{Top-20 reranking performance on the MTEB-multilingual-v2 and MTEB-eng-v2 benchmarks based on the MTEB leaderboard snapshot accessed on August 29, 2026. Only models with complete results on all eight tasks are included. The final average is computed over all eight tasks. Best scores are in bold.}
\label{tab:reranking_mteb_merged_mmteb_eng_complete}
\setlength{\tabcolsep}{3.5pt}
\small
\resizebox{\textwidth}{!}{%
\begin{tabular}{lcrrrrrrrrrrr}
\toprule
\multirow{2}{*}{Model} & \multirow{2}{*}{Size} & \multicolumn{7}{c}{\textit{MMTEB}} & \multicolumn{3}{c}{\textit{MTEB-eng-v2}} & \multirow{2}{*}{Avg.} \\
\cmidrule(lr){3-9} \cmidrule(lr){10-12}
& & Allo & RuBQ & T2 & Voyage & WebLINX & Wiki & Avg. & Ask & Mind & Avg. & \\
\midrule
\textbf{Querit/Querit-Reranker-4B} & 4B & 83.89 & 79.84 & \textbf{69.39} & 70.74 & \textbf{30.81} & 91.78 & \textbf{71.08} & 66.25 & 32.15 & 49.20 & \textbf{65.61} \\
codefuse-ai/F2LLM-v2-14B & 14B & 83.44 & 78.30 & 67.22 & \textbf{83.66} & 19.99 & 90.30 & 70.48 & 67.58 & 33.07 & 50.32 & 65.45 \\
codefuse-ai/F2LLM-v2-8B & 8B & 82.71 & 77.61 & 67.41 & 82.17 & 22.46 & 89.65 & 70.34 & 66.99 & \textbf{33.43} & 50.21 & 65.30 \\
codefuse-ai/F2LLM-v2-4B & 4B & 82.25 & 76.43 & 67.27 & 81.74 & 19.19 & 89.40 & 69.38 & 67.12 & 33.01 & 50.07 & 64.55 \\
ai-sage/Giga-Embeddings-instruct-10B-A1.8B-0826 & 1.8B & 81.06 & \textbf{80.88} & 66.98 & 71.20 & 14.53 & 92.09 & 67.79 & 67.68 & 30.94 & 49.31 & 63.17 \\
ai-sage/Giga-Embeddings-instruct-3B-0826 & 3B & 80.15 & 79.88 & 66.64 & 70.60 & 14.72 & 91.59 & 67.26 & 68.53 & 30.72 & 49.62 & 62.85 \\
codefuse-ai/F2LLM-v2-1.7B & 1.7B & 80.11 & 75.33 & 66.96 & 75.94 & 15.62 & 89.08 & 67.17 & 65.98 & 32.49 & 49.24 & 62.69 \\
Qwen/Qwen3-Embedding-8B & 8B & 84.88 & 73.47 & 68.29 & 65.08 & 11.79 & 90.28 & 65.63 & \textbf{70.20} & 32.93 & \textbf{51.56} & 62.12 \\
Alibaba-NLP/gte-Qwen2-7B-instruct & 7B & 81.10 & 74.13 & 67.80 & 70.06 & 12.50 & 87.73 & 65.55 & 67.58 & 33.36 & 50.47 & 61.78 \\
jinaai/jina-embeddings-v5-text-small & 0.6B & 81.39 & 74.89 & 68.04 & 68.84 & 11.33 & 89.46 & 65.66 & 66.08 & 32.69 & 49.38 & 61.59 \\
jinaai/jina-embeddings-v5-omni-small & 2B & 81.39 & 74.89 & 68.04 & 68.84 & 11.33 & 89.46 & 65.66 & 66.08 & 32.69 & 49.38 & 61.59 \\
Qwen/Qwen3-Embedding-4B & 4B & \textbf{85.13} & 72.28 & 67.27 & 65.61 & 11.30 & 88.89 & 65.08 & 68.81 & 32.71 & 50.76 & 61.50 \\
\textbf{Querit/Querit-Reranker-A0.4B} & 0.4B & 79.39 & 75.37 & 68.98 & 68.59 & 11.63 & 91.05 & 65.84 & 64.02 & 32.52 & 48.27 & 61.44 \\
google/gemini-embedding-001 & N/A & 81.77 & 73.84 & 67.95 & 66.73 & 10.97 & \textbf{92.24} & 65.58 & 64.24 & 32.95 & 48.60 & 61.34 \\
jinaai/jina-embeddings-v5-text-nano & 0.2B & 79.67 & 73.68 & 67.63 & 67.02 & 10.95 & 88.83 & 64.63 & 65.73 & 32.72 & 49.22 & 60.78 \\
jinaai/jina-embeddings-v5-omni-nano & 1B & 79.67 & 73.68 & 67.63 & 67.02 & 10.95 & 88.83 & 64.63 & 65.73 & 32.72 & 49.22 & 60.78 \\
codefuse-ai/F2LLM-v2-0.6B & 0.6B & 78.39 & 71.52 & 67.00 & 72.57 & 13.61 & 87.73 & 65.14 & 64.26 & 31.10 & 47.68 & 60.77 \\
Salesforce/SFR-Embedding-Mistral & 7B & 79.32 & 77.24 & 66.97 & 62.06 & 9.38 & 90.17 & 64.19 & 67.58 & 32.72 & 50.15 & 60.68 \\
Linq-AI-Research/Linq-Embed-Mistral & 7B & 79.82 & 76.84 & 66.89 & 61.24 & 11.28 & 90.13 & 64.37 & 66.82 & 32.06 & 49.44 & 60.64 \\
nvidia/NV-Embed-v1 & 8B & 73.47 & 75.67 & 65.69 & 66.30 & 14.69 & 89.94 & 64.29 & 67.50 & 30.82 & 49.16 & 60.51 \\
% intfloat/e5-mistral-7b-instruct & 7B & 78.32 & 76.32 & 66.90 & 61.62 & 9.51 & 90.26 & 63.82 & 66.98 & 32.59 & 49.78 & 60.31 \\
% nvidia/NV-Embed-v2 & 8B & 73.64 & 77.10 & 66.64 & 63.02 & 11.94 & 90.61 & 63.82 & 67.46 & 31.76 & 49.61 & 60.27 \\
\bottomrule
\end{tabular}
}
\end{table*}

On the combined 8-task MMTEB setting, Querit-Reranker-4B achieves an overall average of 65.61, ranking first among all compared models. Querit-Reranker-A0.4B obtains an overall average of 61.44, remaining competitive with larger small-scale models and validating the effectiveness of the proposed training recipe under a compact active-parameter setting.

\paragraph{Merging strategy comparison.} Table~\ref{tab:merging-comparison-full} provides the task-level results corresponding to Table~\ref{tab:merging-comparison}. SLERP achieves the best score on seven of the eight tasks and the best multilingual, English, and overall averages. TIES-Merging is the strongest non-SLERP alternative, obtaining averages of 65.75, 47.65, and 61.23, but remains below SLERP by 0.09, 0.62, and 0.21 points, respectively. Scoring-head replacement performs best on WebLINX, suggesting that preserving its task-specific head favors this narrow specialization, whereas SLERP more effectively consolidates performance across heterogeneous ranking distributions. These results support the use of SLERP as a training-free method for producing a single deployable reranker without runtime ensembling.
\begin{table*}[t]
\centering
\caption{Comparison of training-free checkpoint merging strategies for Querit-Reranker-A0.4B.}
\label{tab:merging-comparison-full}
\setlength{\tabcolsep}{2pt}
\renewcommand{\arraystretch}{1.05}
\small
\resizebox{\textwidth}{!}{%
\begin{tabular}{lrrrrrrrrrrr}
\toprule
\multirow{2}{*}{Merging strategy}
& \multicolumn{7}{c}{\textit{MTEB-multilingual-v2}}
& \multicolumn{3}{c}{\textit{MTEB-eng-v2}}
& \multirow{2}{*}{Avg.} \\
\cmidrule(lr){2-8}
\cmidrule(lr){9-11}
& Allo & RuBQ & T2 & Voyage & WebLINX & Wiki & Avg. & Ask & Mind & Avg. & \\
\midrule

Linear interpolation
& 79.19
& 75.35
& 68.95
& 67.88
& 11.84
& 90.29
& 65.58
& 63.47
& 30.56
& 47.02
& 60.94 \\

Scoring-head replacement
& 78.98
& 74.46
& 67.88
& 66.38
& \textbf{13.15}
& 89.84
& 65.12
& 62.19
& 30.18
& 46.19
& 60.38 \\

Top-4 layers + head
& 79.23
& 75.29
& 68.93
& 67.67
& 11.96
& 90.25
& 65.56
& 62.46
& 30.77
& 46.62
& 60.82 \\

TIES-Merging
& 79.25
& 75.30
& 68.95
& 68.10
& 12.20
& 90.70
& 65.75
& 63.50
& 31.80
& 47.65
& 61.23 \\

\textbf{SLERP}
& \textbf{79.39}
& \textbf{75.37}
& \textbf{68.98}
& \textbf{68.59}
& 11.63
& \textbf{91.05}
& \textbf{65.84}
& \textbf{64.02}
& \textbf{32.52}
& \textbf{48.27}
& \textbf{61.44} \\

\bottomrule
\end{tabular}%
}
\end{table*}

\paragraph{Stage-wise ablation.} Table~\ref{tab:stage-wise-ablation-full} expands the aggregate results in Table~\ref{tab:stage-wise-ablation} with task-level scores. For A0.4B, Stage I improves all eight tasks and raises the overall average from 26.56 to 56.98, confirming that broad ranking-oriented supervision accounts for most of the gain over the general-purpose backbone. Stage II improves seven of the eight tasks, including a substantial WebLINX improvement from 12.50 to 26.58, and raises the overall average to 60.90. Stage III further improves seven tasks and reaches an overall average of 61.44, although checkpoint consolidation reduces the WebLINX score from 26.58 to 11.63. For 4B, the retrieval-oriented backbone provides a substantially stronger starting point. The gains are therefore distributed more evenly across the three stages: Stage II mainly strengthens the targeted multilingual distributions, particularly Voyage and WebLINX, while Stage III improves six of the eight tasks and raises the multilingual and overall averages from 69.20 and 64.07 to 71.08 and 65.61, respectively.
\begin{table*}[t]
\centering
\caption{Stage-wise ablation of Querit-Reranker. For Stage II, we report the strongest individual pre-merging checkpoint according to the eight-task average.}
\label{tab:stage-wise-ablation-full}
\setlength{\tabcolsep}{2pt}
\small
\resizebox{\textwidth}{!}{%
\begin{tabular}{llrrrrrrrrrrr}
\toprule
\multirow{2}{*}{Model}
& \multirow{2}{*}{Stage}
& \multicolumn{7}{c}{\textit{MTEB-multilingual-v2}}
& \multicolumn{3}{c}{\textit{MTEB-eng-v2}}
& \multirow{2}{*}{Avg.} \\
\cmidrule(lr){3-9}
\cmidrule(lr){10-12}
&
& Allo
& RuBQ
& T2
& Voyage
& WebLINX
& Wiki
& Avg.
& Ask
& Mind
& Avg.
& \\
\midrule

\multirow{4}{*}{\textbf{Querit-Reranker-A0.4B}}
& Base
& 27.38
& 17.44
& 53.18
& 15.96
& 1.65
& 34.58
& 25.03
& 36.97
& 25.29
& 31.13
& 26.56 \\

& Stage I
& 71.16
& 68.86
& 66.51
& 59.79
& 12.50
& 88.16
& 61.16
& 59.66
& 29.16
& 44.41
& 56.98 \\

& Stage II
& 77.75
& 71.06
& 67.42
& 61.19
& 26.58
& 87.87
& 65.31
& 63.50
& 31.81
& 47.66
& 60.90 \\

& \textbf{Stage III}
& 79.39
& 75.37
& 68.98
& 68.59
& 11.63
& 91.05
& \textbf{65.84}
& 64.02
& 32.52
& \textbf{48.27}
& \textbf{61.44} \\

\midrule

\multirow{4}{*}{\textbf{Querit-Reranker-4B}}
& Base
& 85.13
& 72.28
& 67.27
& 65.61
& 11.30
& 88.89
& 65.08
& 68.81
& 32.71
& 50.76
& 61.50 \\

& Stage I
& 82.93
& 79.06
& 69.14
& 70.01
& 13.98
& 91.54
& 67.78
& 65.57
& 32.78
& 49.19
& 63.13 \\

& Stage II
& 82.81
& 78.55
& 69.12
& 71.43
& 21.78
& 91.53
& 69.20
& 64.88
& 32.44
& 48.66
& 64.07 \\

& \textbf{Stage III}
& 83.89
& 79.84
& 69.39
& 70.74
& 30.81
& 91.78
& \textbf{71.08}
& 66.25
& 32.15
& \textbf{49.20}
& \textbf{65.61} \\

\bottomrule
\end{tabular}%
}
\end{table*}

% \section{Example Appendix}
% \label{sec:appendix}

\end{document}